\documentclass[
]{ceurart}

\usepackage{listings}
\usepackage{placeins}
\begin{document}

\copyrightyear{2026}
\copyrightclause{Copyright for this paper by its authors.
  Use permitted under Creative Commons License Attribution 4.0
  International (CC BY 4.0).}

\conference{HAI-Agency: Workshop on Orchestrating Human and AI Agency for Proactive and Reflective Learning, co-located with the 27th International Conference on Artificial Intelligence in Education (AIED 2026), Seoul, Korea, June 27 - July 3, 2026.}

\title{"Help Me, But Don't Track Me": Intervention Timing and Privacy Boundaries for Process-Aware AI Tutors }


\author[1]{Jane Hanqi Li}[%
email=jal221@ucsd.edu,
]
\cormark[1]

\author[1]{Yuhong Zhang}[%
email=yuz291@ucsd.edu,
]

\author[2]{Jiaqi Liu}[%
email=jq-liu23@mails.tsinghua.edu.cn,
]

\author[1]{Tzyy-Ping Jung}[%
email=tpjung@ucsd.edu,
]

\author[1]{Amy Eguchi}[%
email=a2eguchi@ucsd.edu,
]
\cormark[1]


\address[1]{University of California, San Diego, La Jolla, CA, USA}
\address[2]{Tsinghua University, Beijing, China}

\cortext[1]{Corresponding authors: Jane Hanqi Li and Amy Eguchi}





\cortext[1]{Corresponding author.}

\begin{abstract}
As generative AI (GenAI) tools are increasingly used as informal tutors for mathematics learning, future systems may become more proactive and process-aware in deciding when and how to offer support. Yet such support raises an important design tension: help that is timely may also feel interruptive or overly monitoring. To inform the design of process-aware AI tutors, we surveyed 330 secondary school students in China (Grades 7--11) about their preferred tutoring behaviors, attitudes toward proactive intervention, and acceptable use of learning-process data. We found three design-relevant patterns. First, students preferred autonomy-preserving support, such as hints over direct answers. Second, they favored graduated proactive support over constant interruption, preferring small hints first and stronger assistance only as needed. Third, they drew clear privacy boundaries around learning-process data: students were comfortable with problem-solving steps and mistake patterns, but substantially less comfortable with attention- or behavior-related signals. Together, these findings offer early empirical guidance for designing AI tutors that balance timely support with learner agency, and personalization with perceived privacy boundaries in K-12 contexts.
\end{abstract}


\maketitle

\section{Introduction}

Generative AI (GenAI) tools are increasingly used by students as informal tutors for independent mathematics learning \cite{lai2024srl_chatbots,davar2025chatbots_challenges,stando2025aied_math_chatbots}. In many learning contexts, students do not always have access to timely teacher or tutor support, while one-on-one human tutoring remains costly and unevenly available \cite{bloom1984two_sigma,lin2023ai_its_sustainable_sl}. As a result, GenAI is often seen as a scalable way to provide on-demand help when human support is unavailable \cite{lin2023ai_its_sustainable_sl,maity2024genai_personalized_its_arxiv}.

However, effective tutoring is not only about what help is given, but also when and how it is delivered. Prior work in intelligent tutoring systems has long highlighted the challenge of balancing assistance with productive struggle \cite{koedinger2007assistance_dilemma}. Human tutors often use contextual and embodied cues, such as hesitation, confusion, gaze, or disengagement, to decide whether to wait, give a small hint, or step in more directly \cite{sarrafzadeh2008future_its_understand,dmello2012gaze_tutor,khuman2024nonverbal_teaching,hutt2019mind_wandering_classrooms}. As AI tutors become more proactive and process-aware, they may similarly infer when students are stuck and offer support before a request is made \cite{kraus2023roman_proactive_tutoring,maniktala2020avoid_help_avoidance,marwiang2025realtime_feedback_its_math}. Yet this creates a core design tension: support that is more timely may also feel interruptive, autonomy-reducing, or overly monitoring \cite{koedinger2007assistance_dilemma,maniktala2020avoid_help_avoidance,kraus2023roman_proactive_tutoring}.

This tension is especially important in K--12 contexts, where privacy, trust, and learner agency are central concerns \cite{ahn2021codesign_privacy_lak,mintz2023ai_k12_possibilities,paolucci2024la_k12_review}. Process-aware support may rely on learning-process data such as problem-solving steps, repeated mistakes, time stuck, or attention-related signals \cite{regan2019ethical_edtech_bigdata,rehman2025privacy_eye_tracking_ieeeaccess,snoek2025wearables_ethical_classroom}. While such information could make AI support more adaptive, students may not experience all of these data types as equally acceptable, especially when they resemble behavioral monitoring or surveillance \cite{yang2024wpeS_privacy_harms_reviews,shrestha2024youth_privacy_concerns_arxiv}.

To inform the design of process-aware AI tutors, we report survey findings from 330 secondary school students in China. Rather than treating this work as a mature validation study, we position it as early empirical evidence for design. We highlight three design-relevant patterns: students preferred autonomy-preserving scaffolds over direct answers, favored graduated proactive support over constant interruption, and drew sharper privacy boundaries around attention- and behavior-related signals than around problem-solving traces. These findings contribute learner-centered guidance for designing AI tutors that feel supportive rather than intrusive.

We investigate three research questions (RQs):

\textbf{RQ1.} What are students' preferences for GenAI vs.\ human tutors in math learning, and what support behaviors do they prefer when they are stuck or distracted?

\textbf{RQ2.} To what extent do students accept process-aware GenAI support based on learning-process data, and what privacy/autonomy boundaries do they articulate?

\textbf{RQ3.} To what extent do perceptions of proactive intervention (helpful, adaptive, autonomy-preserving, annoying) predict acceptance of process-aware GenAI support?

\section{Related Work}

\subsection{Everyday GenAI Tutoring and Support Timing}

Recent work suggests that students increasingly use GenAI tools as informal tutors during independent study, including in mathematics \cite{lai2024srl_chatbots,davar2025chatbots_challenges,stando2025aied_math_chatbots,rizki2025blended_chatbots_ct_srl}. Unlike traditional intelligent tutoring systems, these tools are often accessed voluntarily and opportunistically, for example when learners feel stuck, confused, or lack immediate support from teachers or peers \cite{peeters2020help_seeking_math,maity2024genai_personalized_its_arxiv}. In such settings, the quality of tutoring depends not only on content quality but also on how the system responds during moments of difficulty.

Prior work on tutoring and help-seeking shows that support can take multiple forms, including direct answers, hints, prompts, and scaffolds, and that these forms differ in how much agency they preserve for learners \cite{wood1999help_seeking_contingent,defeo2017waiting_help_seeking,kennedy1997fears_help_seeking,long2018self_compassion_communication}. Research in ITS has likewise emphasized the ``assistance dilemma''---the challenge of deciding when support should be given and when learners should continue struggling productively \cite{koedinger2007assistance_dilemma}. More recent work on intelligent tutoring and proactive support further suggests that intervention timing and delivery style can shape whether support is experienced as useful or intrusive \cite{maniktala2020avoid_help_avoidance,kraus2023roman_proactive_tutoring,marwiang2025realtime_feedback_its_math}. As AI tutors become more proactive, this question becomes increasingly relevant in everyday GenAI-mediated learning.

\subsection{Process-Aware Support and Privacy Boundaries}

More proactive AI tutoring may depend on learning-process data, such as problem-solving steps, time on task, repeated errors, or even attention-related cues. In principle, these data can help systems infer when a learner is struggling and adapt support more effectively \cite{dmello2012gaze_tutor,hutt2019mind_wandering_classrooms,rehman2025privacy_eye_tracking_ieeeaccess}. However, prior work on educational technology and learning analytics has raised concerns about transparency, trust, privacy, and perceived harm, especially in K--12 contexts \cite{ahn2021codesign_privacy_lak,paolucci2024la_k12_review,beerwinkle2021la_risk_harm_k12,regan2019ethical_edtech_bigdata,mintz2023ai_k12_possibilities}. Recent studies further suggest that students and youth may distinguish between data that are directly tied to learning and data that feel more like surveillance or behavioral monitoring \cite{yang2024wpeS_privacy_harms_reviews,shrestha2024youth_privacy_concerns_arxiv,snoek2025wearables_ethical_classroom}.

Taken together, this literature suggests that the design of future AI tutors must balance adaptivity with learner control. What remains less understood is how students themselves want process-aware AI tutors to behave: what kind of help they prefer, how proactive that help should be, and what kinds of data they view as acceptable in supporting such intervention.

\section{Method}

We conducted a cross-sectional survey to examine how secondary school students think AI tutors should provide support during mathematics learning. The survey was designed to probe design-relevant preferences rather than to validate a comprehensive psychological scale.

\subsection{Participants and Context}
Participants were 330 secondary school students from three public secondary schools in China, operating under the municipal public education system and serving students across lower and upper secondary levels. The sample spanned Grades 7–11: Grade 10 (n=99, 30.0\%), Grade 11 (n=92, 27.9\%), Grade 9 (n=71, 21.5\%), Grade 7 (n=48, 14.5\%), and Grade 8 (n=20, 6.1\%). Students from Grade 12 were invited but participation was minimal due to exam-related constraints. Students were recruited through teacher-mediated invitations. In terms of math study time, the largest group reported spending 30–60 minutes daily (45.8\%), followed by less than 30 minutes (28.2\%), 1–2 hours (17.9\%), and more than 2 hours (8.2\%).

\subsection{Survey Design}
The questionnaire included both closed-ended and open-ended items, adapted from constructs commonly used in prior research on AI-supported learning, academic help-seeking, trust in AI outputs, and learning privacy. The instrument covered four areas: (1) frequency and ways of using GenAI tools for math learning, (2) verification and reliance on GenAI-generated answers, (3) preferences for GenAI versus human tutors and preferred tutoring behaviors when stuck or distracted, and (4) attitudes toward proactive intervention and the acceptable use of learning-process data (e.g., errors, time stuck, steps, attention-related cues).
Most closed-ended items used a 5-point Likert scale (from "Strongly disagree" to "Strongly agree"), along with several multiple-choice questions including select-all-that-apply items. The survey also included open-ended questions asking students to describe: (a) a time GenAI genuinely helped them learn math, (b) what an ideal GenAI tutor should do when they cannot solve a problem, and (c) what a GenAI tutor should never do in a learning context.
Prior to formal data collection, the survey was piloted with a small group of students to check item clarity, reading difficulty, and response time. Based on pilot feedback, item wording and option labels were revised to improve clarity and reduce ambiguity.

\subsection{ Ethics and Procedure}
The survey was voluntary and anonymous. Before starting the questionnaire, students viewed an IRB-approved study information and consent screen that explained the study purpose, procedures, risks, and data handling. Students could opt out by closing the survey at any time, and only those who indicated agreement proceeded to the questionnaire.

\subsection{Data Analysis}
We analyzed closed-ended responses using descriptive statistics (counts, percentages, and means with 95\% confidence intervals where appropriate). Open-ended responses were analyzed using a lightweight thematic analysis. Two researchers first co-developed a shared codebook by independently coding an initial subset of responses and resolving discrepancies through discussion. Using the finalized codebook, three coders coded the full dataset with planned overlap for reliability checks: three overlapping sets of 30 respondents, each double-coded by a different coder pair. The remaining responses were assigned to a single coder. Disagreements in overlapping sets were resolved through discussion, and the codebook was refined as needed. Coding focused on students' expectations for what a GenAI tutor should do, boundary statements about what it should never do, and rationales related to autonomy, annoyance, trust, and privacy.
For RQ3, we fit an ordinal logistic regression model with the 5-level acceptance outcome as the dependent variable and the four perception ratings (adaptive, helpful, annoying, autonomy-related) as predictors, to examine which perceptions of proactive intervention are associated with students' acceptance of process-aware GenAI support.
\section{Results}

We report results organized around three research questions, using a quantitative-led approach complemented by brief qualitative evidence. Analyses are based on N=330 valid survey responses; open-ended themes are summarized as response-level percentages using a three-coder codebook.

\subsection{RQ1: Tutor Preference and Desired Support Behaviors}
\paragraph{Tutor preference.} Most students did not express an exclusive preference for AI tutoring. A majority reported that either a human tutor or a GenAI tutor would be acceptable (54.4\%, 180/330), followed by a preference for a human tutor (42.0\%, 139/330). Only a small minority preferred an AI tutor alone (3.6\%, 12/330).
\paragraph{What "good help" looks like when stuck.} Students overwhelmingly favored autonomy-preserving scaffolds. The most selected option was "Give me some hints so I can continue thinking" (65.6\%, 217/330). Far fewer preferred being given more time to think (17.5\%, 58/330) or being asked what they did not understand (13.9\%, 46/330), and very few wanted the tutor to directly provide the correct answer (3.0\%, 10/330). Open-ended responses reinforced this pattern: over half of students (54.9\%) explicitly stated that an AI tutor should never give the answer directly, and others emphasized preserving space for the student to think (13.1\%).
\paragraph{When intervention is most wanted.} Students most wanted help when they were stuck midway through solving (73.7\%) or could not understand the problem (52.3\%). Fewer wanted intervention when distracted or losing focus (23.0\%), suggesting that attention-related intervention may be more context-sensitive than help for cognitive impasses.
\paragraph{What "distracted" means and how students regain focus. }Open-ended responses clarified that distractions are often triggered by fatigue or low energy (47.6\%), task difficulty or length (46.2\%), and noisy environments or interruptions (41.4\%). Students most often described refocusing through taking a break or resting (55.7\%) and listening to music or background sound (34.3\%), with smaller proportions mentioning help-seeking such as discussing with classmates (15.7\%) or external reminders such as being prompted by a teacher (11.4\%). These self-regulation patterns are relevant for AI tutor design: they suggest that students already have established strategies for managing attention, and that AI intervention during distraction may conflict with rather than complement those strategies.

\subsection{RQ2: Attitudes Toward Proactive Intervention and Data Boundaries}
\paragraph{Attitudes toward proactive support. }Students valued process-aware support but also recognized tradeoffs. On 5-point Likert scales, perceived adaptivity was rated highest (M=4.12, 95\% CI [4.02, 4.22]) and perceived importance of preserving time and autonomy was also high (M=3.94, 95\% CI [3.84, 4.04]). At the same time, proactive support was rated as moderately helpful (M=3.50, 95\% CI [3.38, 3.63]) and moderately annoying (M=2.85, 95\% CI [2.71, 2.99]). This pattern indicates a tension between wanting adaptive help and wanting minimal interruption.
\paragraph{Preferred intervention policy.} When asked what they would prefer if a GenAI tutor could detect that they were stuck, the most common choice was graduated support: providing small hints first and gradually increasing the level of assistance (52.4\%, 173/330). About a quarter preferred help only when requested (24.2\%, 80/330), while 20.3\% (67/330) welcomed proactive reminders or assistance. Only 3.0\% (10/330) rejected proactive intervention entirely.
\paragraph{Overall acceptance of process-aware AI support.} Most students found process-aware help acceptable: 61.3\% (202/330) rated it as "mostly" or "completely acceptable," 30.5\% (101/330) were neutral, and 8.2\% (27/330) found it not acceptable.
\paragraph{Data boundaries. }Students drew clear boundaries around what types of learning-process data a GenAI tutor may use. Willingness was highest for problem-solving steps (74.5\%) and frequent mistake types (73.0\%), moderate for how long they stay stuck on a problem (55.2\%), and substantially lower for whether their attention is focused (34.8\%). A minority of students (8.5\%) indicated they were unwilling to share any of the listed information. Open-ended responses reinforced this pattern, with some students explicitly objecting to AI support that felt intrusive, describing it as invading privacy or monitoring too much.
This tiered pattern suggests that students distinguish between data that support task-level diagnosis, such as what step they are on or what errors they keep making, and data that may feel like monitoring the learner rather than the learning. In other words, many students welcome help, but they do not want that help to come with an expanded sense of being tracked.
\begin{figure}
  \centering
  \includegraphics[width=\linewidth]{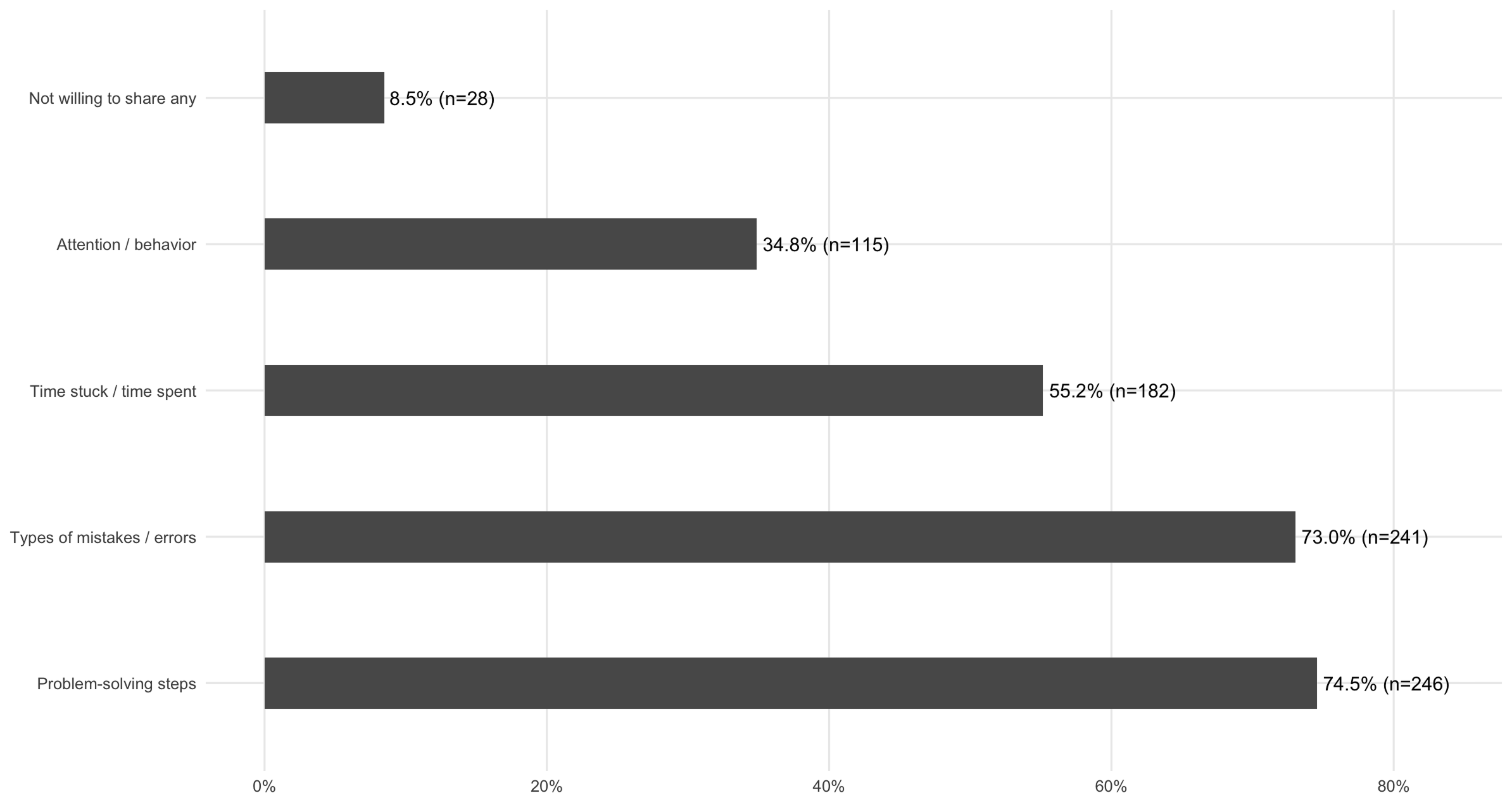}
  \caption{Privacy boundaries: What learning-process data can AI use? (Multiple selection; N=330)}
  \label{fig:privacy_boundaries}
\end{figure}
\FloatBarrier

\subsection{RQ3: Proactive Perceptions Predicting Acceptance of Process-Aware AI Support}
To examine which perceptions of proactive intervention are associated with acceptance, we fit an ordinal logistic regression model with the 5-level acceptance outcome and the four perception ratings (adaptive, helpful, annoying, autonomy-related) as predictors.
Perceived adaptiveness showed the strongest positive association with acceptance (OR > 1; 95\% CI entirely above 1), indicating that students were more likely to endorse higher acceptance levels when they believed proactive support could adjust to their learning state. Perceived helpfulness also exhibited a positive association (OR > 1), though with a comparatively weaker effect. In contrast, perceived annoyance was negatively associated with acceptance (OR < 1; 95\% CI below 1), suggesting that interruption costs meaningfully reduce willingness to accept process-aware intervention.
The autonomy-related perception did not show a clearly distinguishable association with acceptance in the multivariable model (CI crossing 1). This implies that concerns about autonomy may operate through perceived annoyance or overlap with how students evaluate adaptiveness and helpfulness, rather than contributing additional explanatory power once these perceptions are accounted for.

Together, these results suggest that acceptance is shaped by a benefit-cost tradeoff: students are more receptive when process-aware support feels adaptive and useful, but less receptive when it introduces interruption costs. We elaborate on the implications of this pattern in Section 5.2.
\begin{figure} 
  \centering
  \includegraphics[width=\linewidth]{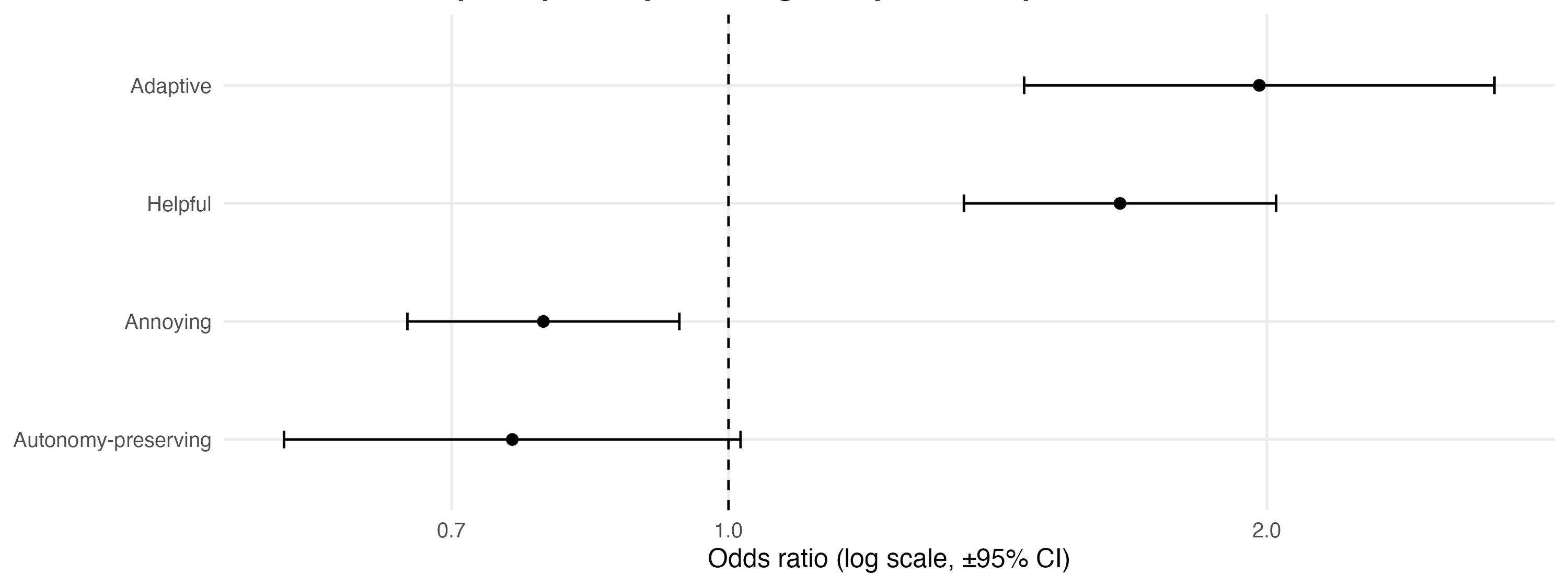}
  \caption{Ordinal logistic regression: Proactive perceptions predicting acceptance of process-aware AI support}
  \label{fig:odds_ratios}
\end{figure}
\FloatBarrier

\section{Discussion}

We organize the discussion around three themes that connect our findings to the HAI-Agency workshop's central question: how to design increasingly autonomous AI systems without diminishing learner agency.

\subsection{Autonomy-First, Low-Interruption Support}

Across RQ1 and RQ2, students consistently preferred an autonomy-preserving model of AI tutoring — one that supports their thinking rather than replacing it. When stuck, they preferred hints that let them continue reasoning and rarely wanted a direct answer. In this study, "good help" was not synonymous with "fast help," but with support that sustains productive struggle and preserves ownership of learning.
At the same time, students were broadly open to proactive intervention, viewing it as potentially adaptive and time-saving, yet they also anticipated interruption costs. The most common preferred intervention policy was a scaffolded strategy: starting with small hints and extending only as needed. This suggests that students favor low-intensity proactive actions that respect their personal workflow rather than aggressive takeover. For the design of agentic AI tutors, this points to a graduated escalation model where AI agency expands only when learner agency appears insufficient, rather than as a default mode of operation.
\\Open-ended responses on distraction further complicate the picture. Students reported that losing focus often stems from fatigue, task difficulty, and environmental factors, and they commonly refocus through self-regulation strategies such as short breaks or background music. This is significant for HAI-Agency because it suggests that attention-related AI intervention risks overriding students' existing self-regulatory capacity. Designing agentic tutors that respect learner agency may therefore require not only calibrating how much help to give, but also recognizing when the learner is already managing the situation themselves.

\subsection{Orchestrating Human and AI Agency: A Benefit-Cost Framing}
The RQ3 regression results offer a mechanism-level account of what shapes acceptance of process-aware AI support. Perceived adaptiveness and helpfulness were positively associated with acceptance, while perceived annoyance was negatively associated. Notably, the autonomy-related item did not show an independent association once the other perceptions were accounted for, suggesting that autonomy concerns may operate through how students evaluate whether interventions feel interruptive or whether adaptiveness is achieved respectfully.
This supports a benefit-cost framing for orchestrating human and AI agency: students are more receptive when AI agency produces tangible benefits (adaptive, well-timed guidance), but less receptive when it introduces interactional costs (irritation, disruption). In practice, this means that acceptance is not simply a function of "more AI awareness", it depends on whether that awareness translates into interventions students experience as appropriately timed, minimally disruptive, and aligned with their own study goals. For the HAI-Agency research agenda, this raises an open design question: how should agentic AI systems signal their awareness without producing the interactional costs that reduce acceptance?

\subsection{Tiered Data Boundaries: "Help Me" Does Not Imply "Track Me"}

The strongest boundary signal in our results is the clear differentiation across types of learning-process data that students are willing to share. Willingness was highest for problem-solving steps and mistake patterns, moderate for time stuck, and substantially lower for attention and behavior signals. This tiered pattern matters because "process-aware" tutoring often bundles multimodal signals under a single label. Our findings argue against treating all learning-process data as equally acceptable.
Students appear to distinguish between data that support task-level diagnosis — what step they are on, what errors they keep making — and data that feel like monitoring the learner rather than the learning. This distinction has direct implications for the HAI-Agency agenda. As AI tutors become more agentic and proactive, the temptation is to maximize the data inputs that feed their decision-making. But our results suggest that learner-centered design requires a data-minimization principle: systems should prioritize lower-sensitivity, task-based signals when adapting support, and treat attention-related signals as requiring clearer justification, transparency, and learner control.
This boundary also connects to broader questions about trust in agentic AI. If students perceive that an AI tutor is collecting more data than they consider appropriate, the resulting discomfort may undermine the very trust that makes proactive support viable. Designing for HAI-Agency, then, not only we need to think about calibrating intervention timing, we also need to consider about calibrating the informational basis on which AI agency operates.

\subsection{Open Questions for HAI-Agency}

Our findings raise several questions for the HAI-Agency community. First, how should escalation policies be co-designed with learners? Our data shows students prefer graduated support, but the specific thresholds and transitions remain underspecified. Second, what does "minimal viable process-awareness" look like? What is the smallest set of data inputs that enables effective proactive support without crossing students' perceived boundaries? Third, how should the orchestration of human and AI agency differ when AI tutoring occurs informally and voluntarily (as in our context) versus in structured classroom settings where teachers also hold agency over the learning process?

\begin{table*}
  \caption{Design-relevant findings and implications for HAI-Agency}
  \label{tab:design_implications}
  \begin{tabular}{p{2.8cm}p{4.2cm}p{4.2cm}p{3.0cm}}
    \toprule
    \textbf{Design challenge} & \textbf{Main finding} & \textbf{Design implication} & \textbf{ Question} \\
    \midrule
    Support when students are stuck & Students preferred hints over direct answers and wanted space to keep thinking. & Start with lightweight scaffolds before escalating to stronger assistance. & When should AI agency expand versus contract during problem solving? \\
    \addlinespace
    Timing of proactive support & Students preferred graduated support over constant interruption or immediate takeover. & Design proactive support as escalation, not interruption. & How should escalation policies be co-designed with learners? \\
    \addlinespace
    Use of learning-process data & Students were more comfortable with steps and mistake patterns than with attention-related signals. & Prioritize low-sensitivity task traces and minimize more intrusive forms of tracking. & What is the minimal data basis for effective proactive support? \\
    \addlinespace
    Predictors of acceptance & Perceived adaptiveness and helpfulness increased acceptance; annoyance reduced it. & Frame process-aware support around tangible benefits while minimizing interactional costs. & How should agentic AI signal awareness without producing disruption? \\
    \bottomrule
  \end{tabular}
\end{table*}
\FloatBarrier

\section{Limitations}

This study has several limitations that should be considered when interpreting the findings.
First, the sample consists of secondary school students from boarding schools in a specific region of China. While this provides a coherent context for examining learner expectations, the findings should not be generalized to non-Chinese populations or to students from other regions of China. Cultural norms around authority, privacy, and technology use in education may shape how students evaluate AI tutoring, and broader demographic and geographic diversity would strengthen the generalizability of these patterns. Future research employing similar methods should examine populations in other regions and countries.
Second, results reflect students' self-reported expectations rather than observed behavior in a live AI tutoring system. Students responded to conceptual descriptions of process-aware support rather than interacting with a concrete interface. Preferences may shift with actual use, for example, students who report discomfort with attention tracking in a survey might respond differently when experiencing a well-designed implementation that provides clear transparency and control. Conversely, features that seem acceptable in the abstract may feel more intrusive in practice. Validating these findings through interactive prototypes that vary intervention policy, interruption style, and transparency mechanisms is an important next step.
Third, privacy-boundary items used a multiple-selection format that captures general willingness across data types but does not address tradeoffs under real design constraints, for instance, when a specific feature requires attention-related signals to function. Students' boundaries may be more or less flexible when tied to concrete benefits or when they can control what is shared. Future work should examine how boundary preferences shift when students are given explicit control over data sharing and can see how specific signals improve the support they receive.
Fourth, the study design is cross-sectional, capturing a snapshot of attitudes at one point in time. As students gain more experience with AI tools and as these tools become more capable, expectations around proactive support and data use are likely to evolve. Longitudinal designs that track how preferences develop with sustained AI tutoring use would complement the present findings.
Finally, our sample is composed entirely of secondary school students, and the findings may not extend to younger learners or to post-secondary contexts where students may have different levels of autonomy, different relationships with technology, and different privacy expectations.
\section{Conclusion}

Despite these limitations, our findings offer learner-centered empirical guidance for the design of process-aware AI tutors, and for the broader HAI-Agency research agenda. Students in our sample generally welcomed timely AI support, especially when stuck, but they preferred autonomy-preserving scaffolds over direct answers and favored graduated proactive assistance that extended only as needed. At the same time, they drew a tiered boundary around what learning-process data a GenAI tutor may use, showing the highest comfort with problem-solving traces, less comfort with time-based indicators, and substantially lower willingness to share attention and behavior signals.
The ordinal regression analysis adds a mechanism-level insight: acceptance of process-aware support is shaped by a benefit-cost tradeoff. Students are more receptive when AI intervention feels adaptive and useful, and less receptive when it feels annoying or disruptive. This suggests that the path to effective agentic AI tutoring runs not through maximizing AI awareness or intervention frequency, but through calibrating both the timing and the informational basis of AI agency to what learners experience as supportive rather than intrusive.
Together, these findings highlight a principle for orchestrating human and AI agency in learning: effective help should preserve learner agency, and personalization should follow a data-minimization approach aligned with learners' perceived boundaries, that is, supporting learners without making them feel tracked.







\section*{Declaration on Generative AI}
During the preparation of this work, the author(s) used ChatGPT and Claude in order to: Grammar checking, proofreading, and formatting revision. After using these tool(s)/service(s), the author(s) reviewed and edited the content as needed and take(s) full responsibility for the publication's content.

\bibliography{sample-ceur}

\appendix



\end{document}